\documentclass[aps,prc,twocolumn,showpacs,a4paper,unsortedaddress,
amsmath,amssymb,byrevtex]{revtex4}

\usepackage[latin1]{inputenc}
\usepackage{graphicx}
\usepackage{dcolumn}
\usepackage{bm}

\begin{document}
\preprint{ffuov/02-01}

\title{An investigation of the role of spectroscopic factors in the breakup reaction of $^{11}\textrm{Be}$}

\author{B. Canbula}\email{bora.canbula@cbu.edu.tr}
\author{R. Bulur}
\author{D. Canbula}
\author{H. Babacan}
\affiliation{Department of Physics, Faculty of Arts and Sciences, Celal Bayar University,
45140, Muradiye, Manisa, Turkey}

\begin{abstract}
The experimental elastic cross section data of the 
projectile $^{11}\textrm{Be}$ on target $^{12}\textrm{C}$ 
at 49.3 MeV/nucleon energy is analysed. The calculations 
for the elastic scattering is performed by the 
phenomenological optical model. The different optical 
potentials to include breakup effects into the 
calculations, which are neutron+$^{12}\textrm{C}$, 
neutron+$^{10}\textrm{Be}$ and $^{10}\textrm{Be}$+$^{12}\textrm{C}$, 
are described with the aid of the global potentials 
for neutron interactions and fitted to experimental 
data for the core and target interaction. 
Also, the first 
analysis of the optical model for $^{10}\textrm{Be}$ on 
target $^{12}\textrm{C}$ at 39.1 MeV is done for 
building the interaction potential of the core and 
the target for $^{11}\textrm{Be}$. For investigating 
the effects of the spectroscopic factor obtained from 
the direct capture process using the nuclear level 
density are compared with the previous cross section and 
spectroscopic factor results. Obtained 
results for the elastic cross section are reproduced the 
experimental data very well, and shows the requirement 
of including spectroscopic properties such as the spectroscopic 
factors and the density of the excited states to explain this elastic 
cross section data.
\end{abstract}


\maketitle

\section{Introduction}\label{introduction}

Experiments with radioactive ion beams (RIB) started a 
new era in the nuclear reaction physics for the last 
decades \cite{tanihata1985a,tanihata1985b,tanihata1988}. 
With regard to probing and understanding the nuclear 
structure, in these experiments, some unexpected properties 
of the nucleus have been discovered. One of the most 
intriguing attributes is the halo structure 
\cite{tanihata1985a}, consisting 
of a core and a weakly-bound valance nucleon(s) to 
this core. Up to now, this phenomenon has been greatly 
investigated experimentally on the various targets 
\cite{pires2011,kolata1992,cortinagil1997} and caused 
a challenge for nuclear reaction theoreticians to 
reproduce the experimental data \cite{pires2011,thompson1993,johnson1997}.  

$^{11}\textrm{Be}$ is one of the four one-neutron 
halo nuclei together with $^{19}\textrm{C}$ \cite{nakamura1999}, 
and newly reported $^{31}\textrm{Ne}$ \cite{nakamura2014}, 
and $^{37}\textrm{Mg}$ \cite{kobayashi2014}. Some experiments 
has been conducted for understanding of structure of 
$^{11}\textrm{Be}$. Firstly, 
Tanihata et. al \cite{tanihata1988} observed a large 
radii for $^{11}\textrm{Be}$ compared to $^{10}\textrm{Be}$ 
in cross section measurements with targets at 90 MeV/A, 
and concluded the halo structure for $^{11}\textrm{Be}$ 
originating from its small neutron separation energy of 
0.503 MeV. A few years later, Fukuda et al. \cite{fukuda1991} 
confirmed this conclusion in elastic scattering of 
$^{11}\textrm{Be}$ on C and Al targets at 33 MeV/nucleon. 
Since these distinguished works, $^{11}\textrm{Be}$ has been 
still studied experimentally \cite{cortinagil1997,lapoux2008,dipietro2010,anne1993,fukuda2004} 
and theoretically \cite{summers2006a,howell2005,dasso1999}. 

One experimental study about
$^{11}\textrm{Be}$ is performed 
by Cortina-Gil et al. \cite{cortinagil1997} for the cross 
section of the elastic scattering 
on $^{12}\textrm{C}$ at 49.3 MeV/nucleon incident energy. 
The first theoretical investigation of this measurement 
is an adiabatic approximation assuming 
no internal motion between the valance nucleon and the core in 
projectile, and also neglecting the interactions between the valance 
nucleon and the target nucleus \cite{johnson1997}. 
Also, in the same year, Al-Khalili et al. \cite{alkhalili1997} 
investigated the same reaction with the few-body Glauber model, 
in which the particles of the projectile are considered 
as following straight line paths through the interaction 
field of the target. In addition to these studies, the 
continuum-discretized coupled-channels (CDCC) method was 
applied to this elastic scattering by Takashina et. al. 
\cite{takashina2003}, and also they used the same parameter 
set for the optical potentials between the projectile 
components and the target as in Ref. \cite{alkhalili1997}. 
In this non-adiabatic method, due to the very low neutron 
or proton separation energy, the continuum states of the projectile 
above this threshold energy are discretized to finite number of 
states using momentum bins. Including the breakup effects 
into the theoretical calculations of mentioned methods gives 
almost the same results. 

In the present study, the elastic scattering of the projectile 
halo nucleus $^{11}\textrm{Be}$ on the target $^{12}\textrm{C}$ 
at 49.3 MeV/nucleon \cite{cortinagil1997} is investigated as a 
breakup reaction using the optical model with the aid of a nuclear 
structure model. Different from other studies, 
the optical model potential used for the interaction 
between the core nucleus $^{10}\textrm{Be}$ and the target 
$^{12}\textrm{C}$ is obtained by fitting to elastic cross section 
data at 39.1 MeV/nucleon. This data \cite{lapoux2008} is investigated 
with optical model for the first time in this study. As for the interaction 
between the valance neutron of the halo nucleus and the target, 
the optical potential is deduced from an interpolation for different 
incident energies of neutron on $^{12}\textrm{C}$ target by means of 
global potential of Ref. \cite{koning2003}. 
In order to describe non-elastic contributions, 
we use a surface potential, which can be named as DPP (dynamical 
polarization potential) or VCP (virtual coupling potential), in 
our calculations. Finally, a binding potential is employed for 
n+$^{10}\textrm{Be}$ system. Unlike similar studies, we  
determine the value of the spectroscopic factor, describing 
the wave function of $^{11}\textrm{Be}$ in terms of the 
wave function of  $^{10}\textrm{Be}$, with the method given in Ref. 
\cite{goriely1997} for the direct neutron capture reaction  
$^{10}\textrm{Be}$+$n \rightarrow {^{11}\textrm{Be}}$+$\gamma$. 
However, we use a new nuclear level density (NLD) model 
\cite{canbula2014}, which strongly depends on the deformation 
of nucleus. 

This paper organized as follows: The method used in this study is 
presented in Sec. \ref{theory}, the results obtained by this method 
are given for the breakup reaction of $^{11}\textrm{Be}$ in Sec. 
\ref{results}, and finally in Sec. \ref{conclusions}, concluding 
remarks drawn from this study are given.

\section{Theory}\label{theory}

Since the mid-fifties, the optical model has been greatly 
used to investigate the elastic scattering cross section for 
both light and heavy ions in a wide range of incident energy. 
The optical model considers the projectile and 
the target nuclei as the structureless particles in order 
to avoid many-body problems in nuclear physics calculations, 
and describes the interaction between the projectile and target 
with an effective potential. 
In this work, since we included breakup effects, the halo projectile $^{11}\textrm{Be}$ 
is considered as a two-body system, which consists of $^{10}\textrm{Be}$ 
core and a valance neutron. Therefore, we define the 
effective potentials between projectile components and the target 
$^{12}\textrm{C}$, which are n+$^{12}\textrm{C}$, 
$^{10}\textrm{Be}$+$^{12}\textrm{C}$ and n+$^{10}\textrm{Be}$ 
as used Ref. \cite{summers2006b} 
\begin{equation}\label{eq:effectivepotential}
{U_\mathrm{eff}}={U_\mathrm{CT}}+{U_\mathrm{VT}}+{U_\mathrm{CV}}, 
\end{equation} 
where C, T, V correspond to $^{10}\textrm{Be}$ core, $^{12}\textrm{C}$ 
target and valance nucleon, respectively. An effective potential is a  combination of 
the following terms as 
\begin{equation}\label{eq:effectivepotentialterms}
{U(r)}={V_{l}}(r)+{V_\mathrm{Coul}}(r)+{V_\mathrm{Vol}}(r)+{V_\mathrm{Sur}}(r)+V_{\mathrm{SO}}(r).
\end{equation} 
The first term is the centrifugal potential, which is defined as 
\begin{equation}\label{centrifugalpotential}
V_{l}(r)=\frac{\mathrm{\hbar} \mathrm{l}(\mathrm{l}+1)}{2\mathrm{m}\mathrm{r}^2}
\end{equation} 
traditionally. Uniformly 
charged sphere assumption is employed for nucleus, 
\begin{equation}\label{eq:coulombpotential}
V_{C}(r) = \left\lbrace
\begin{array}{ll}
\displaystyle{\frac{Z_{P} Z_{T} \mathrm{e}^{2}}{2 R_{C}} \left ( 3 - 
\frac{r^{2}}{{R_{C}}^{2}} \right )} & r \le R_{C} \\
\displaystyle{\frac{Z_{P} Z_{T} \mathrm{e}^{2}}{r}} &  r \ge R_{C}
\end{array}
\right.
\end{equation}
where the charge radius $R_c$ is defined as ${R_c}={r_c({A_P^{1/3}}+{A_T^{1/3}})}$, the Coulomb potential 
parameter ${r_c}$ is taken as 1.2 fm in this work. 
In the optical model, the volume term in a effective potential has a crucial role and 
can be described with the real part of this term. However, 
for inelastic contributions, 
an imaginary part is added to the volume term for the purpose 
of considering absorption of the reaction flux from the 
elastic channel to inelastic reaction channels. 
Therefore, 
conventionally the volume term consists of real 
and imaginary parts in reaction studies, 
\begin{equation}\label{eq:volumepotential}
V_{\mathrm{Vol}}(r) = \frac{- V_0}{1+\mathrm{exp} \left( \frac{r-R_{\mathrm{v}}}{a_{\mathrm{v}}} \right)} + \frac{- i W_0}{1+\mathrm{exp} \left( \frac{r-R_{\mathrm{w}}}{a_{\mathrm{w}}} \right)}
\end{equation}
where potential 
depths, radius and surface diffuseness parameters for both real 
and imaginary parts should be adjusted elastic scattering data. 
Even if the investigated reaction is the elastic 
scattering, non-elastic contributions can still exist in the elastic channels. 
To include these contributions, the surface potential 
are used 
\begin{equation}\label{eq:surfacepotential}
V_{\mathrm{Sur}}(r) = \frac{-4 V_0 \mathrm{exp}\left( \frac{r-R_{\mathrm{v}}}{a_{\mathrm{v}}} \right)}{1+\mathrm{exp} \left( \frac{r-R_{\mathrm{v}}}{a_{\mathrm{v}}} \right)} + \frac{-4i W_0 \mathrm{exp}\left( \frac{r-R_{\mathrm{w}}}{a_{\mathrm{w}}} \right)}{1+\mathrm{exp} \left( \frac{r-R_{\mathrm{w}}}{a_{\mathrm{w}}} \right)},
\end{equation}
which is in derivative form of the volume term. Final term in Eq. 
\eqref{eq:effectivepotentialterms} is the spin-orbit potential
\begin{equation}\label{eq:spinorbitpotential}
V_{\mathrm{SO}}(r)={\left( \frac{\hbar}{m_{\pi}c} \right)}^{2} \frac{1}{r} \frac{\mathrm{d}}{\mathrm{d}r} \frac{V_{\mathrm{SO}}}{1+\mathrm{exp} \left( \frac{r-R_{\mathrm{SO}}}{a_{\mathrm{SO}}} \right)} 2 \mathbf{L} \cdot \mathbf{s}
\end{equation}
where $(\hbar/m_{\pi}c)^{2}=2 \, \mathrm{fm}^{2}$.

The optical potential parameters in these equations 
can be determined from elastic scattering data. 
As a first step in fitting procedure of potential 
parameters, the geometrical parameters are adjusted 
to positions of peaks occurred in data. 
Afterwards, the potential depths of all used optical model potentials are fitted to experimental 
data to give the minimum $\chi^2$ value. 

In the case of halo nucleus $^{11}\textrm{Be}$, 
the spectroscopic factor as a structure
property is used to describe the ground state and the first excited 
state of $^{11}\textrm{Be}$ in terms of $^{10}\textrm{Be}$. 
The spectroscopic factor can be determined from 
the fitting to experimental cross section data of transfer 
or direct capture processes, and also they can be obtained 
theoretically from the shell model calculations. In 
literature, many transfer processes include the spectroscopic 
factor value of $^{11}\textrm{Be}$ for 
$^{9}\textrm{Be}$(t,p)$^{11}\textrm{Be}$ 
\cite{pullen1962,robertson1978,selove1978,liu1990}, 
$^{10}\textrm{Be}$(d,p)$^{11}\textrm{Be}$ 
\cite{goosman1970,auton1970,zwieglinski1979,schmitt2013} and 
$^{11}\textrm{Be}$(p,d)$^{10}\textrm{Be}$ \cite{winfield2001} 
reactions. However, the experimental data of the direct 
capture cross section for $^{10}\textrm{Be}$(n,$\gamma$)$^{11}\textrm{Be}$
is not available, but the direct capture cross section data can be 
deduced from Coulomb dissociation \cite{mengoni1997}. As a tool 
for calculations of the light ion cross sections 
such as direct capture processes, the nuclear level 
density has a crucial role to reproduce 
the measured data and define the spectroscopic factor. Therefore, the relation 
between the direct capture cross section and the nuclear level 
density, which is the number of the excited levels around an excitation
energy, can be defined as \cite{goriely1997}
\begin{equation}\label{directcapture}
\sigma^{DC}(E)=<S> \int_{0}^{Sn} \sum_{J_{f},\Pi_{f}} \rho(E_{f},J_{f},\Pi_{f}) \sigma_{f}^{cont}(E)dE_{f}, 
\end{equation} 
where S represents the average spectroscopic factor, and $\rho$ is 
the level density function in terms of the excitation energy 
$E_f$, total angular momentum $J_f$, and 
the parity $\Pi_f$ of compound nucleus. In the present work, 
we calculate the direct neutron capture cross section 
and compare to deduced data \cite{mengoni1997} from Coulomb 
dissociation of $^{11}\textrm{Be}$ measured by Nakamura et. al. 
\cite{nakamura1994}. To do this calculation, Laplace-like formula \cite{canbula2014} is used for 
the energy dependence of the nuclear level density parameter in Fermi gas model. 
According to this formula the level density parameter strongly 
depends on the deformation of the nucleus, and the results obtained 
with this formula are very successful to describe low-lying 
collective levels compared to other phenomenological 
level density models \cite{koning2008}. Therefore, keeping in 
mind that $^{10}\textrm{Be}$ and $^{11}\textrm{Be}$ 
are well-deformed nuclei, we expect that this formula is convenient 
to explain the neutron capture cross section data of $^{10}\textrm{Be}$. 
In the following section, we 
will give the optical potential parameters, which are used in this study, 
and the results of our calculations.

\section{Results and Discussion}\label{results}
\begin{figure}[t!]
\includegraphics[width=\columnwidth]{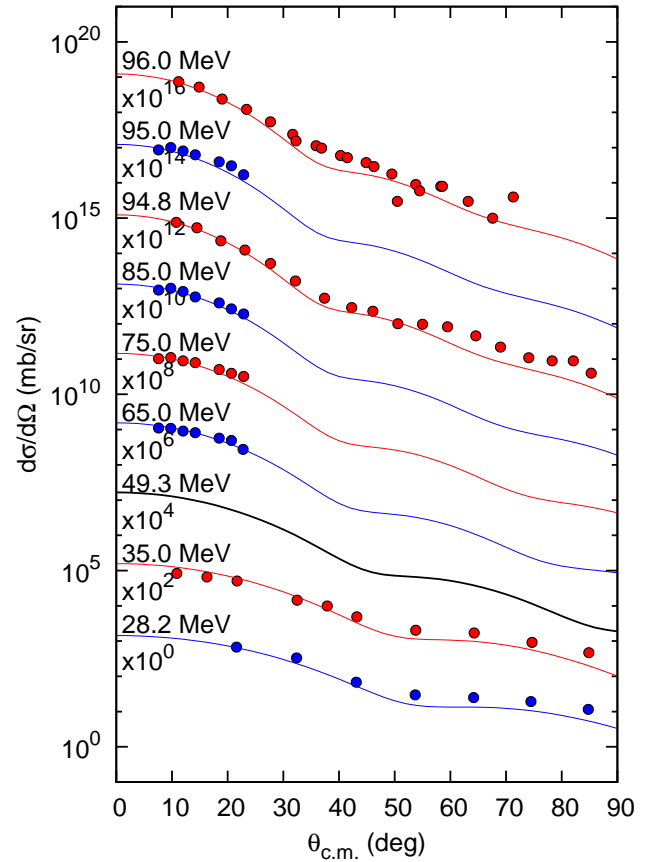}
\caption{\label{Fig1}(Color online) Cross sections for n+$^{12}\textrm{C}$ at 
28.2 MeV \cite{yamanouti1989}, 35.0 MeV \cite{niizeki1990}, 
65.0 MeV \cite{osborne2004}, 75.0 MeV \cite{osborne2004}, 85.0 
MeV \cite{osborne2004}, 94.8 MeV \cite{mermod2006}, 95.0 MeV 
\cite{osborne2004}, 96.0 MeV \cite{klug2003}. Obtained results 
using the optical potentials for 49.3 MeV incident 
energy are represented by black line.}
\end{figure}

To describe the interactions between the projectile and target, 
we consider the weakly-bound nucleus $^{11}\textrm{Be}$ 
as $^{10}\textrm{Be}$+n on $^{12}\textrm{C}$ target. For this 
purpose, firstly we focus on the interaction between the 
neutron and the target. Great number of experimental data in 0-100 MeV energy range \cite{yamanouti1989,niizeki1990,osborne2004,mermod2006,klug2003} is found for the elastic 
scattering of neutron on $^{12}\textrm{C}$, and can be used 
to define the effective potential between the valance nucleon and the target 
for in case $^{11}\textrm{Be}$. Unfortunately, for 49.3 MeV incident energy, no 
experimental data is available. Thus, an interpolation of the global parametrization \cite{koning2003} is used. The results obtained with this global 
potential are given in Figure \ref{Fig1}. As seen from figure, this interpolation of the 
global parametrization for n+$^{12}\textrm{C}$ at 49.3 MeV incident energy gives 
us a good agreement in wide range energy. 

\begin{figure}[b!]
\includegraphics[width=\columnwidth]{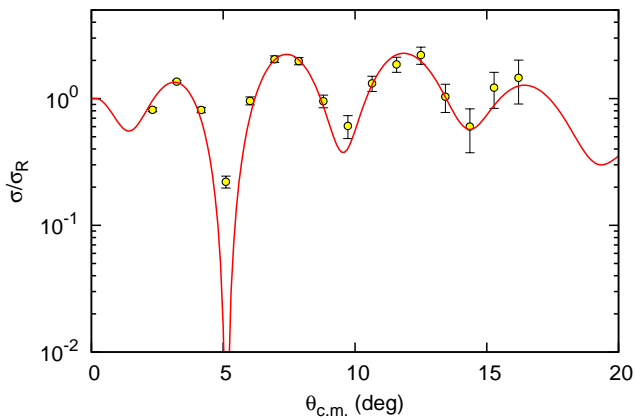}
\caption{\label{Fig2}(Color online) Cross sections for $^{10}\textrm{Be}$+$^{12}\textrm{C}$ target at 39.1 MeV. Solid red line represents 
the results obtained by using the optical potential parameters given in Table \ref{Tab01}. The experimental data is 
taken from \cite{lapoux2008}.}
\end{figure}

In contrast to $^{11}\textrm{Be}$, very long-lived ($\mathrm{T_{1/2}=1.5 \times 10^6}$y) 
and a tightly-bound 
nucleus $^{10}\textrm{Be}$ has a greater neutron separation 
energy, which is 6.81 MeV. One of the experimental study about 
$^{10}\textrm{Be}$ is Lapoux et. al. \cite{lapoux2008} in which they 
measured the elastic cross section for $^{10}\textrm{Be}$ and 
$^{11}\textrm{Be}$ projectiles on proton and $^{12}\textrm{C}$ 
targets at 39.1 MeV/nucleon and 38.4 MeV/nucleon, respectively, 
and these data was investigated using the microscopic 
Jeukenne-Lejeune-Mahaux nucleon-nucleus potential for proton 
target and framework of folding model for C target.
Unlike the other studies, \cite{alkhalili1997,takashina2003} in 
order to be more physical and reliable, the potential parameters
describing the interaction between the core and the target 
are adjusted to the elastic scattering data at 39.1 MeV/nucleon 
energy \cite{lapoux2008}. 
Our obtained values of the potential 
depth parameters are shallow compared to their potential. We use the experimental $\beta_{2}$ quadrupole 
deformation value, which is -0.6 \cite{vermeer1983}, for the first
$(2^+)$ excited level of $^{12}\textrm{C}$, which is 4.4 MeV. 
Also, in order to take into account the non-elastic contributions 
caused from the interactions at surface region, 
additionally one can add the surface term into the effective 
potential. This potential is sometimes referred as surface 
term, derivative of Woods-Saxon potential form, 
DPP and VCP, and can be obtained by different methods. The parameters of DPP 
can be obtained from microscopical \cite{hussein1994,khoa1995,pacheco2000} or 
phenomenological \cite{canto1991,canto1992,yabana1992a,yabana1992b,takigawa1992,alkahlili1995,sakuragi1995} 
calculations by fitting to the experimental data. For $^{10}\textrm{Be}$+$^{12}\textrm{C}$, we used a
phenomenological DPP obtained from the fit process to the experimental 
data combined with a volume term. Obtained results for this elastic scattering and the optical potential parameters used in this calculation are 
given in \ref{Fig2} and Table \ref{Tab01}, respectively. Except from the well-known phenomenon at $5^{\circ}$, the data is reproduced well.

\begin{table}[t!]
\caption{Adjusted potential parameters for n+$^{12}\textrm{C}$, 
n+$^{10}\textrm{Be}$, $^{10}\textrm{Be}$+$^{12}\textrm{C}$ and 
$^{11}\textrm{Be}$+$^{12}\textrm{C}$ interactions. $r_c$ is taken 
as 1.20 fm for the Coulomb interaction.}\label{Tab01}
\begin{ruledtabular}
\begin{tabular}{llccccc}
{Interaction} & {Type} &{$\mathrm{V_0}$(MeV)} & {$\mathrm{r_v}$(fm)} & {$\mathrm{a_v}$(fm)} \\
{Potential}   & {} &{$\mathrm{W_0}$(MeV)} & {$\mathrm{r_w}$(fm)} & {$\mathrm{a_w}$(fm)} \\
\hline \\ [-1.8ex]
{n+$^{12}\textrm{C}$} &{Volume}&{37.5} &{1.127}&{0.676}&\\
{}            &{} &{4.90} &{1.127}&{0.676}&\\
{}                    &{Surface} &{0.00} &{1.306}&{0.543}&\\
{}            &{} &{4.15} &{1.306}&{0.543}&\\
{}                   &{Spin-Orbit}   &{4.68} &{0.903}&{0.590}&\\
{}         &{} &{-0.39}&{0.903}&{0.590}&\\
\hline \\ [-1.8ex]
{$^{10}\textrm{Be}$+$^{12}\textrm{C}$} &{Volume}&{15.049} &{0.950}&{0.580}&\\
{}                             &{} &{23.326} &{1.100}&{0.630}&\\
\hline \\ [-1.8ex]
{n+$^{10}\textrm{Be}$} &{Volume}&{37.5} &{1.127}&{0.676}&\\
\hline \\ [-1.8ex]
{$^{11}\textrm{Be}$+$^{12}\textrm{C}$} &{Surface}&{42.793} &{0.950}&{0.580}&\\
{SF=1.48}                    &{}&{3.935} &{1.100}&{0.630}&\\
\hline \\ [-1.8ex]
{$^{11}\textrm{Be}$+$^{12}\textrm{C}$} &{Surface}&{29.635} &{1.100}&{0.580}&\\
{SF=0.71,0.62}               &{}&{1.036} &{1.100}&{0.630}&\\
\end{tabular}
\end{ruledtabular}
\end{table}

Many authors analysed the elastic scattering of the 
halo-nuclues $^{11}\textrm{Be}$ on target $^{12}\textrm{C}$ for this incident energy by 
different theoretical models \cite{johnson1997,alkhalili1997,takashina2003}. 
However, none of these studies is incorporated the 
nuclear structure to explain the data. 
On the other hand, adding the nuclear structural information into the 
reaction calculations for such a weakly-bound system 
as the halo nucleus is played a crucial role to contribute the 
agreement between the predictions and the experimental data. 
The spectroscopic factor as a nuclear structure property is one of the most important 
ingredients for the theoretical cross section calculations 
of both light and heavy ions. 
There are many methods which can be used for determining 
the value of the spectroscopic factor. Of course, the easiest 
method is to fit the spectroscopic factor values to the experimental 
cross section data directly, but the traditional way of estimating 
spectroscopic factor is to use the shell model in which the 
spectroscopic factor is defined as the square of normalization 
of the overlap integral between the wave function of the valance 
nucleon in the state of the target nucleus and the residual nucleus. 
Also, the spectroscopic factor is a key ingredient for the direct 
capture process for which the related cross section 
often dominates the total cross section at the very low energies 
of astrophysical interest. The direct capture process is can be used for 
obtaining the spectroscopic factor and known to play a notable 
role for light exotic nuclei systems for which few, or even no 
resonant states are available. Although many works containing 
the spectroscopic factors derived from the transfer processes 
are existing for the halo nucleus $^{11}\textrm{Be}$, 
the direct neutron capture cross section data for $^{10}\textrm{Be}$ 
to compose $^{11}\textrm{Be}$ is not available in the literature. 
However, the direct capture cross sections can be obtained deduced 
data from the Coulomb dissociation. 

\begin{figure}[t!]
\includegraphics[width=\columnwidth]{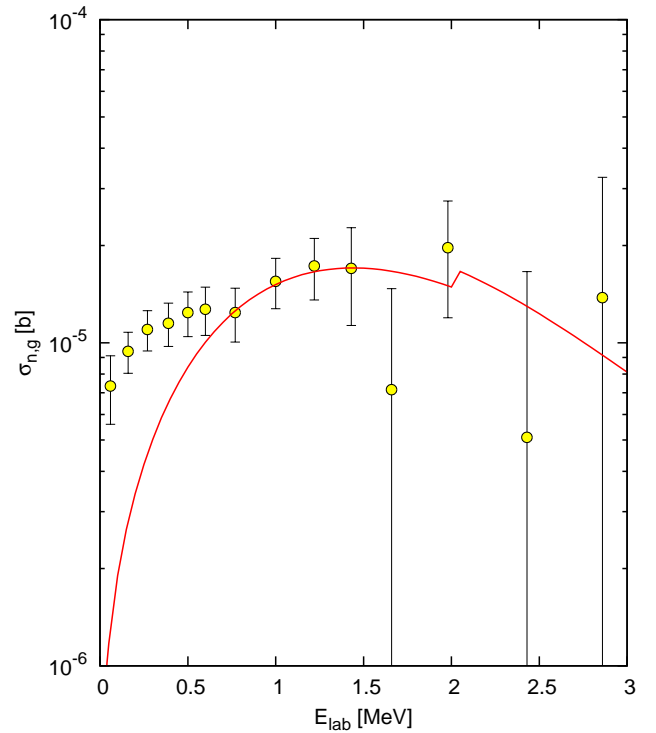}
\caption{\label{Fig3}(Color online) The direct neutron capture cross section 
results for $^{10}\textrm{Be}$(n+$\gamma$)$^{11}\textrm{Be}$ 
reaction at 0-3.MeV lab. energy. The solid red line represents 
the results of the present work using the level density 
model \cite{canbula2014}, and obtained spectroscopic factor 
value is 1.48. 
The deduced experimental data from Coulomb dissociation data of $^{11}\textrm{Be}$ \cite{nakamura1994} is taken from 
Ref. \cite{mengoni1997}.}
\end{figure} 

In obtaining the spectroscopic factor with the aid of the direct capture cross 
section calculations, the most important component is the nuclear level density. 
Generally, the reasons for not trusting to level 
density models to use in such calculations are their insufficient 
agreements with the experimental observables and their way 
of taking into account the collective effects.
For overcoming these challenges, recently, we introduced a new 
Laplace-like formula \cite{canbula2014} for the NLD parameter to 
improve the predictive power for describing the low-lying 
collective levels, which are well known to be of vital importance 
for the direct capture process. With this formula, a great agreement 
is achieved with the experimental observables. 
Therefore, the direct neutron capture cross section calculation based on this level 
density model for $^{10}\textrm{Be}$(n+$\gamma$)$^{11}\textrm{Be}$ 
processes is shown in Figure \ref{Fig3}. 
Although the data could not be reproduced below 0.5 MeV, in the rest of the energy range 
the same behaviour is well explained. 
The obtained average value for the spectroscopic factor is 1.48. 
The value of parameters used in our level density calculation are 
1.345 for $\tilde{a}$ and 0.285 for $\beta$, which are taken from our previous study.
 
\begin{figure}[t!]
\includegraphics[width=\columnwidth]{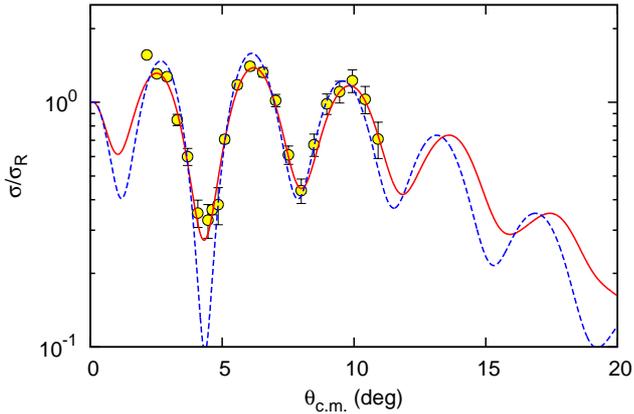}
\caption{\label{Fig4}(Color online) The elastic cross section results 
for $^{11}\textrm{Be}$+$^{12}\textrm{C}$ 
at 43.9 MeV. The dashed blue line is the breakup calculation with the spectroscopic factors 
0.71 for the ground state and 0.62 for the first excited state  
\cite{schmitt2012}. The solid red line represents the cross section 
result with the spectroscopic factor value of 1.48 obtained from 
the direct capture cross section. The experimental data 
is taken from \cite{cortinagil1997}.}
\end{figure}

Considering $^{11}\textrm{Be}$ as a two-body projectile, 
all values of the optical potential parameters are given 
in Table \ref{Tab01} for n+$^{12}\textrm{C}$, 
n+$^{10}\textrm{Be}$ and $^{10}\textrm{Be}$+$^{12}\textrm{C}$ 
interactions. The parameter values of potential depths for 
$^{10}\textrm{Be}$+$^{12}\textrm{C}$ at 39.1 MeV incident energy 
are rearranged as 46.3 MeV and 13.8 MeV 
of real and imaginary parts, respectively. The same procedure 
is repeated for the surface potential as 9.820 MeV and 3.661 MeV. 
Also, to include 
non-elastic contribution for $^{11}\textrm{Be}$+$^{12}\textrm{C}$, 
a surface potential is added to effective potential. 
Moreover, to compare our results, we perform another calculation with the spectroscopic factor obtained 
by Schmitt et al. \cite{schmitt2012} which is 0.71 for the ground state and 0.62 for the first excited state, respectively. The results of 
this calculation is also shown in Figure \ref{Fig4}
with dashed blue line. In our calculations 
the average value of the spectroscopic factor is taken as the spectroscopic factor of the ground state. 
Since the spectroscopic factor of the first excited state has less effect on the results, the value of this factor is taken as 1.0. Finally, our prediction for the elastic 
scattering cross section of $^{11}\textrm{Be}$ on $^{12}\textrm{C}$ is 
shown in Figure \ref{Fig4} with solid red line.
As seen from this figure, the inclusion of the nuclear level density with the 
Laplace-like formula in the reaction calculations has a very well effect on  
reproducing the cross section data. Also, the fit method we used 
for the optical potential parameters, which is to 
adjust the geometrical parameters to positions of peaks and the depths 
to give minimum $\chi^2$, effects the agreement in a positive way.

\section{Conclusions}\label{conclusions}
In summary, we have investigated the elastic scattering 
cross section data of the projectile $^{11}\textrm{Be}$ on $^{12}\textrm{C}$ 
target at 49.3 MeV/nucleon \cite{cortinagil1997}. To include breakup 
effects into the calculations, the different optical 
potentials for n+$^{12}\textrm{C}$, n+$^{10}\textrm{Be}$ 
and $^{10}\textrm{Be}$+$^{12}\textrm{C}$ are described. 
Also, the present study contains 
the first analysis of the phenomenological optical model 
for 39.1 MeV incident energy of the projectile $^{10}\textrm{Be}$ 
on $^{12}\textrm{C}$ target.
Obtained results are in better agreement 
compared to the microscopic study of Lapoux et al \cite{lapoux2008}, 
which is the first and the only study of this reaction. 

Not only the effect of including the spectroscopic factor into 
calculations are found significant for the breakup reaction 
of $^{11}\textrm{Be}$ but also adjusting the geometrical parameters 
to positions of peaks and the depths to give minimum $\chi^2$ 
gives positives contributions to reproduce the scattering data.

The theoretical framework used for obtaining the spectroscopic 
factor by using the nuclear nuclear level density to calculate 
the direct neutron capture cross section is employed for the 
first time in breakup reaction calculation of $^{11}\textrm{Be}$. 
Moreover, the nuclear level density is used for the first time 
as a spectroscopic tool in a light exotic nuclei induced reaction. 
Consequently, beside the success of the nuclear level density with 
the Laplace-like formula for the level density parameter 
\cite{canbula2014} as a structure model, the results show that this 
new method seems appropriate to perform the reaction calculations.

\begin{acknowledgments}
This work was supported by the Turkish Science and Research 
Council (T\"{U}B\.{I}TAK) under Grant No. 112T566. Bora Canbula 
acknowledges the support through T\"{U}B\.{I}TAK PhD Program 
fellowship B\.{I}DEB-2211 Grant.
\end{acknowledgments}

\end{document}